\documentclass[twocolumn,showpacs,aps,prd,superscriptaddress,floatfix]{revtex4}
\usepackage{graphicx}
\usepackage{verbatim}
\usepackage{dcolumn}
\usepackage{bm}
\usepackage{amsmath}
\usepackage{epsfig}

\newcommand{\BaBarYear}       {10}
\newcommand{\BaBarNumber}     {007}
\newcommand{\SLACPubNumber} {14116}

\newcommand{\BaBarType}      {PUB}  

\input babarsym.tex

\def\bcount    {\ensuremath { 471 \times 10^{6}} }

\def\btodgam {\ensuremath {b\to d\gamma} }
\def\btosgam {\ensuremath {b\to s\gamma} }

\def\VtdVts	{\ensuremath{|V_{td} / V_{ts}|}}
\def\Btorhoomgam {\ensuremath{B \to (\rho,\omega)\gamma}}
\def\BtoKgam  {\ensuremath{B\to K^*\gamma}}
\def\BtoXsgam  {\ensuremath{B\to X_s\gamma}}
\def\BtoXdgam  {\ensuremath{B\to X_d\gamma}}

\def\itypeone 	{\ensuremath{\Bz \to \pipi \gamma}}
\def\itypetwo	{\ensuremath{\Bp \to \pip \piz \gamma}}
\def\itypethree	{\ensuremath{\Bp \to \pipi \pip \gamma}}
\def\itypefour	{\ensuremath{\Bz \to \pipi \piz \gamma}}
\def\itypefive	{\ensuremath{\Bz \to \pipi \pipi \gamma}}
\def\itypesix	{\ensuremath{\Bp \to \pipi \pip \piz \gamma}}
\def\itypeseven	{\ensuremath{\Bp \to \pip \eta \gamma}}
\def\itypesone 	{\ensuremath{\Bz \to \Kp \pim \gamma}}
\def\itypestwo 	{\ensuremath{\Bp \to \Kp \piz \gamma}}
\def\itypesthree{\ensuremath{\Bp \to \Kp \pipi \gamma}}
\def\itypesfour {\ensuremath{\Bz \to \Kp \pim \piz \gamma}}
\def\itypesfive	{\ensuremath{\Bz \to \Kp \pim \pipi \gamma}}
\def\itypessix 	{\ensuremath{\Bp \to \Kp \pim \pip \piz \gamma}}
\def\itypesseven {\ensuremath{\Bp \to \Kp \eta \gamma}}

\def\figurebox#1#2#3{%
    \def\arg{#3}%
   \ifx\arg\empty
    {\hfill\vbox{\hsize#2\hrule\hbox to #2{\vrule\hfill\vbox to #1{\hsize#2\vfill}\vrule}\hrule}\hfill}%
    \else
    {\hfill\epsfbox{#3}\hfill}%
    \fi}

\begin{document}

\pagestyle{empty}

\begin{flushleft}
\babar-\BaBarType-\BaBarYear/\BaBarNumber \\
SLAC-PUB-\SLACPubNumber\\
\end{flushleft}

\title{

  {\large \bf \boldmath
Study of $B \to X \gamma$ Decays and Determination of $|V_{td}/V_{ts}|$ 
}
}
%
\author{P.~del~Amo~Sanchez}
\author{J.~P.~Lees}
\author{V.~Poireau}
\author{E.~Prencipe}
\author{V.~Tisserand}
\affiliation{Laboratoire d'Annecy-le-Vieux de Physique des Particules (LAPP), Universit\'e de Savoie, CNRS/IN2P3,  F-74941 Annecy-Le-Vieux, France}
\author{J.~Garra~Tico}
\author{E.~Grauges}
\affiliation{Universitat de Barcelona, Facultat de Fisica, Departament ECM, E-08028 Barcelona, Spain }
\author{M.~Martinelli$^{ab}$}
\author{A.~Palano$^{ab}$ }
\author{M.~Pappagallo$^{ab}$ }
\affiliation{INFN Sezione di Bari$^{a}$; Dipartimento di Fisica, Universit\`a di Bari$^{b}$, I-70126 Bari, Italy }
% ...
\collaboration{The \babar\ Collaboration}
\noaffiliation

\begin{abstract}
\noindent
Using a sample of 471 million \BB\ events collected with the \babar\ detector, we study 
the sum of seven exclusive final states $B \to X_{s(d)}\gamma$, where $X_{s(d)}$ is a strange (non-strange) hadronic system 
with a mass of up to 2.0\gevcc. 
After correcting for unobserved decay modes, we obtain a branching fraction for
\btodgam\ of
($9.2 \pm 2.0(stat.)\pm 2.3(syst.))\times 10^{-6}$ in this mass range,  and a branching fraction for \btosgam\ of
($23.0 \pm 0.8 (stat.)\pm 3.0(syst.))\times 10^{-5}$ in the same mass range.
We find $\frac{{\cal{B}}(\btodgam)}{{\cal{B}}(\btosgam)} = 0.040 \pm 0.009 (stat.) \pm 0.010(syst.),$
from which we determine $|V_{td}/V_{ts}|=0.199 \pm 0.022(stat.) \pm 0.024(syst.) \pm 0.002(th.)$.

\end{abstract}


\maketitle


The decays \btodgam\ and \btosgam\  are flavor-changing neutral current processes forbidden 
at tree level in the Standard Model (SM). The leading-order processes are one-loop electroweak 
penguin diagrams, for which the top quark is the dominant virtual particle. 
In theories beyond the SM, new virtual particles may appear in the loop, 
which could lead to measurable effects on experimental observables such as 
branching fractions and $CP$ asymmetries~\cite{bsm}. 
In the SM the inclusive rate for \btodgam\ is suppressed relative to \btosgam\ by a factor 
$|V_{td}/V_{ts}|^2$, where $V_{td}$ and $V_{ts}$ are Cabibbo-Kobayashi-Maskawa matrix elements. 
Measurements of $|V_{td}/V_{ts}|$ using the exclusive modes 
$\Btorhoomgam$ and $\BtoKgam$~\cite{bellerhog,babarrhog} are now 
well-established, with 
theoretical uncertainties of 7\% from weak annihilation and
hadronic form factors~\cite{BJZ}. 
This ratio can also be obtained from the $B_d$ and $B_s$ 
mixing frequencies~\cite{bsmixing}. It is important to confirm the consistency
of the two methods of determining $|V_{td}/V_{ts}|$, since new physics effects 
would enter in different ways in mixing and radiative decays.  
A measurement of the branching fractions of inclusive $\btodgam$ relative to $\btosgam$ would 
determine $|V_{td}/V_{ts}|$ with reduced theoretical uncertainties
compared to that from exclusive modes~\cite{AAG}.

This letter supersedes of~\cite{old}, and presents the first significant observation of 
the $\btodgam$ transition in the hadronic mass range $M(X_d)>1.0\gevcc$,
resulting in a significant improvement in the determination
of $|V_{td}/V_{ts}|$ via the ratio of inclusive widths. 
Inclusive $\btosgam$ and $\btodgam$ rates are extrapolated from the 
measurements of the partial decay rates to seven exclusive final states
(see Table~\ref{tab:itypes}) in the hadronic mass ranges  
$0.5<M(X_d)<1.0\gevcc$ (low mass, containing the previously measured $K^*$, $\rho$ and $\omega$ resonances) and $1.0<M(X_d)<2.0\gevcc$ (high mass). 
We combine these measurements and make a model-dependent extrapolation to higher 
hadronic mass to obtain an 
inclusive branching fraction (${\cal B})$ for $b\to (s,d)\gamma$.
These measurements use the full dataset of \bcount\ \BB\ pairs collected at 
the \FourS\ resonance
at the PEP-II B factory with  the \babar\ detector~\cite{babar}.

\begin{table}
\centering
\caption{\label{tab:itypes} The reconstructed decay modes. Charge conjugate states 
are implied throughout this paper.}
\vspace*{2mm}
\begin{tabular}{ll} \hline \hline
\BtoXdgam\ 	& \BtoXsgam\  \\ \hline
\itypeone\ 	&\itypesone\\
\itypetwo\ 	&\itypestwo\\
\itypethree\ 	&\itypesthree\\
\itypefour\ 	&\itypesfour \\
\itypefive\ 	&\itypesfive \\
\itypesix\ 	&\itypessix \\
\itypeseven\ 	&\itypesseven\\
\hline \hline
\end{tabular}
\end{table}


High energy photons are reconstructed from an isolated energy cluster in the
barrel of the calorimeter, with shape consistent with a single photon, and energy
$1.15 < E_\gamma^* < 3.50 \gev$, where $^*$ denotes the center-of-mass (CM) frame. 
We remove photons that can form a \piz($\eta$) candidate in association with 
another photon of energy greater than 30 (250)\mev if the 
two-photon invariant mass is in the range 
$110<m_{\gamma\gamma}<160\  (520<m_{\gamma\gamma}<560)\mevcc$ 
for the low mass region and 
$95<m_{\gamma\gamma}<155\ (530<m_{\gamma\gamma}<565) \mevcc$ 
for the high mass region. 

Charged pion and kaon candidates are selected from well-reconstructed tracks. 
We use a pion selection algorithm to differentiate pions from kaons, 
with a typical 
selection efficiency of 95\% and kaon mis-identification rate of 4\%. 
Kaons are identified as tracks failing the pion selection criteria. 
We reconstruct $\piz(\eta$) candidates from pairs of photons
of minimum energy 20 \mev\ with an invariant mass 
$115<m_{\gamma\gamma}<150$ $(470<m_{\gamma\gamma}<620) \mevcc$. 
We require all pion, $\eta$ and kaon candidates 
to have a momentum in the laboratory frame greater than 600 (425)\mevc\ in the 
low (high) mass region. 

The selected pion, $\eta$, kaon and high-energy photon candidates are combined 
to form \B\ meson candidates consistent with one of the seven 
decay modes. 
The charged particles are combined to form a common vertex with a $\chi^2$  
probability greater than 1\%. 
We use the kinematic variables $\de = E^*_B - E^*_{\rm beam}$, 
where $E^*_B$ is the energy of the $B$ meson 
candidate and $E^*_{\rm beam}$ is the beam energy, 
and  $\mes = \sqrt{ E^{*2}_{\rm beam}-{\vec{p}}_{B}^{\;*2}}$, 
where ${\vec{p}}_B^{\;*}$ is the momentum of the $B$ candidate. 
We consider candidates in the range $-0.3 < \de <0.2 \gev$ 
and $\mes >  5.22\gevcc$. 

Contributions from continuum processes
($e^+e^- \to q \qbar$, with $q = u, d, s, c$) are reduced by 
considering only events for which the ratio $R_2$ of second-to-zeroth 
order Fox-Wolfram moments~\cite{fox} is less than 0.98.  
To further discriminate between the jet-like continuum background and 
the more spherically symmetric signal events, 
we compute the angle $\theta_T^*$ between the 
photon momentum and the thrust axis of the rest of the event (ROE) 
and require $|\cos(\theta_T^*)|<0.8$. The ROE is defined as all 
charged tracks and neutral energy deposits 
that are not used to reconstruct the $B$ candidate.

Ten other event shape variables that distinguish between 
signal and continuum events are combined in a neural network (NN). These
include the ratio $R'_2$, which is $R_2$ is calculated in the frame recoiling
against the photon momentum, the $B$ meson production angle with respect to 
the beam axis in the CM frame, $\theta_B^*$, 
and the L-moments~\cite{legendre} of the ROE with respect to either
the thrust axis of the ROE or the direction of the high energy photon.
Differences in lepton, pion and kaon
production between background and \B\ decays are exploited by including several
flavor-tagging
variables applied to the ROE~\cite{babartag}.
Using the NN output, we reject 99\% of continuum background 
while preserving 25\% of signal decays

After all selections are applied, there remain events with more than one $B$ candidate. 
In these events the candidate with the reconstructed 
$\piz$ or $\eta$ mass closest to nominal is retained. 
Where there is no $\piz$ or $\eta$ we retain the candidate with the highest vertex 
$\chi^2$ probability.


The signal yields in the data for the sum of the seven decay modes 
are determined from two-dimensional 
extended maximum likelihood fits to the $\de$ and $\mes$ distributions. 
We consider the following contributions: signal, combinatorial backgrounds 
from continuum processes, backgrounds from other $B$ decays, 
and cross-feed from mis-reconstructed  $B\to X\gamma$ decays.  
The fits to \BtoXdgam\  events contain components from misidentified  
\btosgam\ decays, and 
we neglect the small \btodgam\  background in 
the fits to \BtoXsgam\ events. 

Each contribution is modeled by a probability density function (PDF) 
that is determined from Monte Carlo (MC) simulated events unless otherwise specified. 
For the misidentified signal cross-feed components, we use 
a binned two-dimensional PDF to account for correlations. 
All the other PDFs are products of one-dimensional functions of $\de$ and $\mes$. 
For signal, the $\mes$ spectrum is described by a Crystal 
Ball function~\cite{CryBall}, and $\de$ by a 
Cruijff function~\cite{cruijff}. 
The parameters of these functions are determined from the fit to the 
high-statistics \BtoXsgam\ data sample. We use these fitted values 
to fix the signal shape in the fits to \BtoXdgam\ events. 

The remaining $B$ backgrounds contain a small 
component that peaks in \mes\ but not \de, 
which is modeled by a Gaussian distribution in \mes. 
Continuum and other non-peaking backgrounds are described by an ARGUS 
shape~\cite{argus} in \mes  
and a second-order polynomial in \de. 

\begin{figure}[bht]
\begin{center}
\includegraphics[width=3.4in]{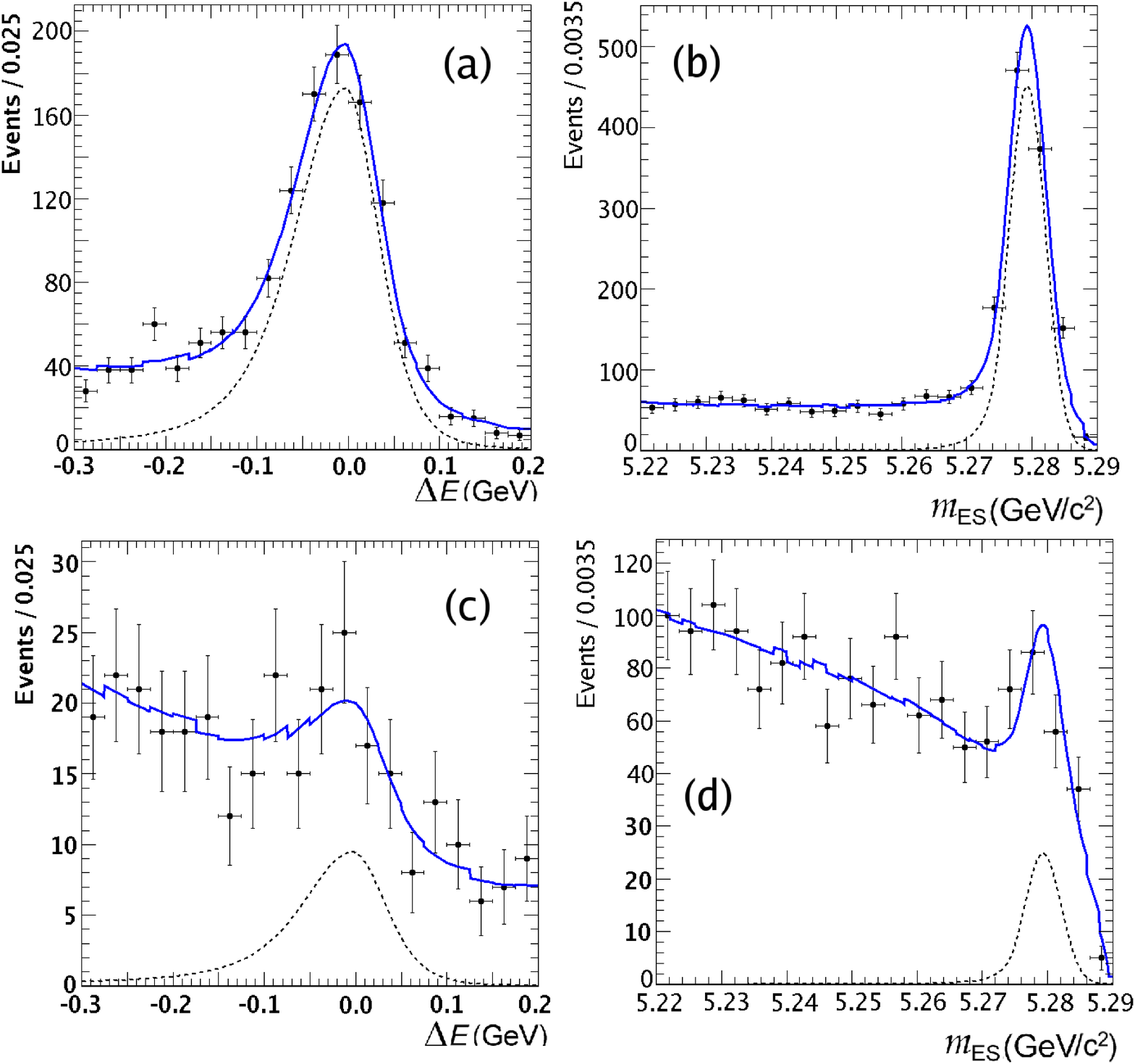}
\caption{Projections of $\de$ with $5.275<\mes<5.286\gevcc$ for
(a) \BtoXsgam\ and (b) \BtoXdgam, and of 
$\mes$ with $-0.1<\de<0.05\gev$ for
(c) \BtoXsgam\ and (d) \BtoXdgam\ in the mass range 1.0-2.0\gevcc.
Data points are compared with the sum of all the fit contributions 
(solid line). The jagged line is an artifact of the fit projection over the 
sum of several binned histograms. 
The dashed line shows the signal component. 
}
\label{fig:proj} 
\end{center}
\end{figure}

We perform separate fits for \BtoXdgam\ and \BtoXsgam\ in
each of the hadronic mass ranges 0.5-1.0\gevcc and 1.0-2.0\gevcc.
For each of the four fits, we combine the component PDFs
and fit for the signal, generic $B$ and continuum yields, the ARGUS 
and two polynomial shape parameters. 
We scale the cross-feed contributions proportionally to the 
fitted signal yield, re-fit and iterate until the procedure converges. 
Projections of \mes\ and \de\ from fits to data for \BtoXsgam\ and 
\BtoXdgam\ are shown in the
high mass regions in Figure~\ref{fig:proj}. 
Table~\ref{tab:bfs} gives the signal yields, efficiencies 
(after corrections for systematic effects) 
and partial branching fractions (${\cal PB}$). 
We calculate ${\cal PB}$ using 
${\cal PB}(B\to X\gamma) = N_S / (2\ \epsilon  N_{B\bar{B}})$, 
where $N_{B\bar{B}}$ is the number of $B\bar{B}$ pairs 
in the data sample.

\begin{table*}
\centering
\caption{\label{tab:bfs}
Signal yields ($N_S$), efficiencies ($\epsilon$), 
partial branching fractions ($\cal PB$), inclusive branching 
fractions ($\cal B$) and the ratio of inclusive branching fractions 
for the measured decay modes. 
The first error is statistical and second is systematic (including an 
error from extrapolation to missing decay modes, for the inclusive $\cal{B}$).}
\begin{tabular}{ccccccc} \hline \hline
			&\ $M(X_s)0.5-1.0$\ 	&\ $M(X_d) 0.5-1.0$\ 	&\ $M(X_s) 1.0-2.0$\ 	&\ $M(X_d)1.0-2.0$\    	&\ $M(X_s)0.5-2.0$\ 	&\ $M(X_d)0.5-2.0$\  \\ \hline
$N_S$	 		&$804\pm 33$\ 		&  $35\pm 9$\		& $990\pm 42$\  	& $56\pm 14$\ 		& - 			& - \\ 
$\epsilon$		&  4.5\%		& 3.1\%			& 1.6\%			& 1.9\%			& - 			& - \\ 
${\cal {PB}}(\times 10^{-6})$ & $19\pm 1\pm 1$ 	& $1.2\pm 0.3\pm 0.1$ 	& $66\pm 3\pm 6$ 	& $3.2\pm 0.8\pm 0.5$ 	& - 			& - \\ 
${\cal{B}}(\times 10^{-6})$ & $38\pm 2\pm 2$ & $1.3\pm 0.3\pm 0.1$ 	& $192\pm 8\pm 29$ 	& $7.9\pm 2.0\pm 2.2$ & $230\pm 8\pm 30$ 	& $9.2\pm 2.0\pm 2.3$ \\ \hline

$\frac{{\cal{B}}(b\to d\gamma)}{{\cal{B}}(b\to s\gamma)}$ & \multicolumn{2}{c}{$0.033\pm 0.009\pm 0.003$}  & \multicolumn{2}{c}{-} & \multicolumn{2}{c}{$0.040\pm 0.009 \pm 0.010$} \\ \hline \hline
\end{tabular}
\end{table*}


We have investigated a number of sources of systematic uncertainty 
in the measurement of the partial branching fractions,
some of which are common to both \BtoXdgam\ and \BtoXsgam\, and cancel in the 
ratio of branching fractions (see Table~\ref{tab:syst}: those that do not cancel in the ratio are marked by an asterisk). Uncertainties in tracking efficiency, 
particle identification, \g\ and \piz\ reconstruction, and the \piz/$\eta$ veto  
have been evaluated using independent control samples 
of data and MC simulated events, and incorporated into our analysis. 
Uncertainty due to the NN selection has been evaluated by 
comparing the efficiency of the selection in data and MC for the  \BtoXsgam\ 
events, which are relatively free of background, 
assuming that potential discrepancies between data and MC are the same for the \BtoXdgam\ sample. 
The means and widths of the signal PDF are varied within the range allowed 
by the fit to the \BtoXsgam\ data, accounting for correlations. 
Other PDF parameters are also varied within 
the 1$\sigma$ limits determined from the fit to MC. 
We vary the \btosgam\ background in the 
fit to \BtoXdgam\ by the statistical uncertainty on our measurement of those decays. 
The signal cross-feed originating from our measured channels is varied by the statistical 
uncertainty on our measurement; other signal cross-feed backgrounds by $\pm$50\%. 
An additional uncertainty on the efficiency arises from the  
fragmentation of the hadronic system among the measured final states. 
For \BtoXsgam\, the uncertainty is constrained by the errors on the
measured data; 
for \BtoXdgam\  an estimate is obtained from 
the difference between the default phase-space fragmentation (see below) and 
a re-weighting using the measured data/MC differences in \BtoXsgam.

\begin{table}
\centering
\caption{\label{tab:syst}
Systematic errors on the measured partial and inclusive branching fractions ${\cal B}$.
Systematic errors that do not cancel in the ratio of rates are marked with (*). }
\vspace*{2mm}
\begin{tabular}{lcccc} \hline\hline
Systematic 		& \multicolumn{2}{c}{$M(X_s)$} 	& \multicolumn{2}{c}{$M(X_d)$} \\
Error Source 		& 0.5-1.0& 1.0-2.0 		& 0.5-1.0 & 1.0-2.0 \\ \hline

Track selection         & 0.3\%     & 0.4\%    & 0.3\%	& 0.4\% \\ 
Photon reconstruction   & 1.8\%     & 1.8\%    & 1.8\%    & 1.8\% \\ 
\piz/$\eta$ reconstruction  & 0.9\% & 1.1\%    & 1.4\%    & 1.6\% \\ 
Neural network          & 1.1\%     & 4.9\%    & 1.1\%    & 4.9\% \\ 
\B\ counting            & 0.6\%     & 0.6\%    & 0.6\%    & 0.6\% \\ 

PID (*)                 & 2.0\%     & 2.0\%    & 2.0\%    & 2.0\% \\ 
Fit bias (*)            & 0.1\%     & 0.9\%    & 4.9\%    & 6.5\% \\ 
PDF shapes (*)          & 2.3\%     & 0.6\%    & 3.7\%    & 3.4\% \\ 
Histogram binning (*)   & 0.8\%     & 0.2\%    & 1.8\%    & 1.8\% \\ 
Background (*)          & 0.8\%     & 1.2\%    & 5.9\%    & 7.0\% \\ 
Fragmentation (*)       & -              & 3.3\%    & -             &  5.1\% \\
Signal model            & -              & 5.8\%    & -             &  6.0\% \\ \hline 
Error on partial ${\cal B}$& 4.0\%    & 9.0\%    & 9.3\%    & 14.2\% \\ \hline
Missing $\ge 5$ body 	& 		 &  9.6\%  	&   & 18.2\% \\
Other missing states 	& 		 &  7.5\% 	 &  &  15.3\% \\
Spectrum Model 		& 		 &  1.8\% 	 &  &  1.6\%  \\ \hline
Error on inclusive ${\cal B}$ 	&  4.0\% &  15.2\%  &  9.3\%  &  27.7\% \\ \hline\hline
\end{tabular}
\end{table}


To obtain inclusive ${\cal B}(\btosgam)$ and ${\cal B}(\btodgam)$ we need to 
correct the partial $\cal B$ values in Table~\ref{tab:bfs} 
for the fractions of missing final states. 
After correcting for the 50\% of missing decay modes with neutral kaons,
the low mass \BtoXsgam\ measurement is found to be consistent
with previous measurements of the rate for \BtoKgam~\cite{hfag}.
For the low mass \BtoXdgam\ region, we correct for the small amount 
of non-reconstructed $\omega$ final states (for example, 
$\omega \to \piz \gamma$),  and find a partial
branching fraction consistent with previous measurements of 
$\BR(\Btorhoomgam)$~\cite{hfag}. 
We assume that non-resonant decays do not contribute in this region. 

In the high mass region, 
the missing fractions depend on the fragmentation of the 
hadronic system and are expected to be different for $X_d$ and $X_s$.
In our signal MC, fragmentation is modeled
by selecting an array of final-state particles and resonances according
to the phase-space probability of the final state, as implemented by
JETSET~\cite{JETSET}. 
We further constrain the distribution of $X_s$ final states to that observed
for our seven decay modes as well as the distributions of a number of other states
measured in~\cite{xsgam}. 
According to this MC model we reconstruct 43\%\ of the total width in \btodgam, and 36\%\ 
in \btosgam.  
A further 37\%\ of the width of \btosgam\ is constrained by the isospin relation 
between charged and neutral kaon decays. 
We explore the uncertainty in the correction for missing modes by
considering several alternative models: replacing 50\% of $\btosgam$
and $\btodgam$ hadronic final states with a mix of resonances; varying
$\btosgam$ fragmentation constraints within their statistical uncertainties;
and setting the $\btodgam$ fragmentation rates to those of their 
corresponding $\btosgam$ states. 
The resulting missing fractions vary by up
to 50(40)\% relative to the nominal model in \BtoXsgam(\BtoXdgam).
We therefore independently vary final states with $\ge 5$ stable hadrons, or
with $\ge 2\pi^0$ or $\eta$ mesons,
by $\pm$50(40)\%. 

Results for the corrected $\cal B$ values are shown in 
Table~\ref{tab:bfs}. Adding the two mass regions, taking into account a
partial cancellation of the missing fraction uncertainties in the ratio
of \btodgam\ to \btosgam, we find
${\cal B}(\btodgam) / {\cal B}(\btosgam) = 0.040 \pm 0.009(stat.) \pm 0.010(syst.) $	
in the mass range $M(X)<2.0\gevcc$. 	

	
We correct for the unmeasured region $M(X) > 2.0$ \gevcc
using the 
spectral shape from Kagan-Neubert~\cite{kaganneubert} 
with the kinetic parameters $(m_b,\mu_{\pi}^2) = 
(4.65 \pm 0.05, -0.52 \pm 0.08)$ 
extracted from fits of $b\to s\gamma$ and 
$b\to c\ell\nu$ data~\cite{OliverHenning},
yielding corrections of $1.66 \pm 0.03$;   
the spectra for \btosgam\ and \btodgam\ are expected to be almost identical. 

Conversion of the ratio of inclusive branching fractions to
the ratio $|V_{td}/V_{ts}|$ is done according to~\cite{AAG}, 
which requires the Wolfenstein parameters $\bar{\rho}$ and $\bar{\eta}$ as input. 
However, since the world average of these quantities relies on previous 
measurements of \VtdVts\, we instead re-express $\bar{\rho}$ and $\bar{\eta}$
 in terms of the world average of the independent CKM angle $\beta$~\cite{hfag}. 
This procedure yields a value of 
$ |V_{td}/V_{ts}| = 0.199 \pm 0.022(stat.) \pm 0.024(syst.) \pm 0.002(th.)$, 
compatible and competitive with more model-dependent determinations
from the measurement of the exclusive modes
$\Btorhoomgam$ and $\BtoKgam$~\cite{bellerhog,babarrhog}.                     

In summary, we have measured the inclusive $\btosgam$
and $\btodgam$ transition rates using a sum of seven
final states in the hadronic mass range up to 
2.0\gevcc, making the first significant observation of
the $\btodgam$ transition in the region above 1.0\gevcc.
The value of $|V_{td}/V_{ts}|$ derived 
from these measurements has an experimental uncertainty 
approaching that from the measurement of exclusive
decays $\Btorhoomgam$ and $\BtoKgam$, but a significantly
smaller theoretical uncertainty.

\section{Acknowledgments}
\label{sec:Acknowledgments}

\vspace{-0.2cm}
We are grateful for the excellent luminosity and machine conditions
provided by our \pep2\ colleagues, 
and for the substantial dedicated effort from
the computing organizations that support \babar.
The collaborating institutions wish to thank 
SLAC for its support and kind hospitality. 
This work is supported by
DOE
and NSF (USA),
NSERC (Canada),
CEA and
CNRS-IN2P3
(France),
BMBF and DFG
(Germany),
INFN (Italy),
FOM (The Netherlands),
NFR (Norway),
MES (Russia),
MICIIN (Spain),
STFC (United Kingdom). 
Individuals have received support from the
Marie Curie EIF (European Union),
the A.~P.~Sloan Foundation (USA)
and the Binational Science Foundation (USA-Israel).


\begin{thebibliography}{99}
\bibitem{legendre}
L-moments are defined as
$L_{i} \equiv \sum_{j} p^{*}_{j}
\cdot|\cos{\theta^{*}_{j}}|^{i}/\sum_{j} p^{*}_{j}$ and
$\tilde{L}_{i} \equiv \sum_{j} p^{*}_{j}
\cdot|\sin{\theta^{*}_{j}}|^{i}/\sum_{j} p^{*}_{j}$, where
$p^{*}_j$ and $\theta^{*}_{j}$ are the momentum and angle with
respect to a given axis, respectively, for each particle $j$ in the ROE.
\bibitem{cruijff}
The Cruijff function is a centered Gaussian with different 
left-right resolutions and non-Gaussian tails:
$f(x) = \exp\left((x-m)^2 / (2\sigma^2_{L,R} + \alpha_{L,R}(x-m)^2) \right)$.
\end{thebibliography}
\end{document}